# Spatiotemporal Photonic Emulator of Potential-free Schrödinger Equation

Qian Cao[1,2,3,†], Nianjia Zhang[1,†], Andy Chong[4,5,*], and Qiwen Zhan[1,2,3,6,7,*]

[1] School of Optical-Electrical and Computer Engineering, University of Shanghai for Science and Technology, 200093 Shanghai, China

[2] Zhangjiang Laboratory, Shanghai, China

[3] University of Shanghai for Science and Technology, Shanghai Key Laboratory of Modern Optical System, Shanghai, China

[4] Pusan National University, Department of Physics, Busan, Republic of Korea

[5] Institute for Future Earth, Pusan National University, Busan 46241, Republic of Korea

[6] Westlake Institute for Optoelectronics, Fuyang, Hangzhou 311421, China

[7] International Institute for Sustainability with Knotted Chiral Meta Matter (WPI-SKCM2), Hiroshima University, Higashihiroshima, Hiroshima 739-8526, Japan

[†] These authors contribute equally to this work.

[*] Corresponding authors: chong0422@pusan.ac.kr, qwzhan@usst.edu.cn

**Abstract**: Photonic quantum emulator utilizes photons to emulate the quantum physical behavior of a complex quantum system. Recent study in spatiotemporal optics has enriched the toolbox for designing and manipulating complex spatiotemporal optical wavepackets, bringing new opportunities in building such quantum emulators. In this work, we demonstrate a new type of photonic quantum emulator enabled by spatiotemporal localized wavepackets with spherical harmonic symmetry. The spatiotemporal field distribution of these wavepackets has the same distributions of the wavefunction solutions to the potential-free Schrödinger equation with two controllable quantum numbers. A series of such localized wavepackets are experimentally generated with their localized feature verified. These localized wavepackets can propagate invariantly in space-time like particles, forming a new type of photonic quantum emulator that may provide new insight in studying quantum physics and open up new applications in studying light-matter interactions and quantum optics.

Introduction

Four decades ago, Feynman proposed the idea that the operation of a complex quantum system can be simulated by another simpler and more feasible system[1]. This has led to the creation of so-called quantum emulator that uses physical or chemical process in a non-quantum system such as atoms[2,3], trapped ions[4-10], and even nuclear magnetic resonance[8,9] to emulate the quantum physical behaviors in quantum mechanics. Among these quantum emulators, photonic quantum emulator[10] has drawn great attentions and have become a promising platform for studying quantum physics due to several reasons such as the Bosonic nature of photon, the accessibility for complex light fields, and the possibility for creating entangled states for photons.

One type of photonic quantum emulator is built upon utilizing the angular momentum property of photon, more specifically, the orbital angular momentum (OAM) state of that photon that is related with the spiral phase structure of the light field[11] and the spin angular momentum (SAM) state of photon that is related with the polarization state of the light[12-14]. Facilitated by photonic OAM and SAM states, scientists have succeeded in emulating quantum walk[15], finding eigen-state solution in quantum chemistry[16], and creating artificial Bell states[17-19] using light as photonic quantum emulator.

Above examples on using photonic OAM for building photonic quantum emulators are all related with the longitudinal OAM, meaning the direction of OAM is in parallel with the propagation direction of the light. Recently, photonic transverse OAM has been theoretically and experimentally confirmed in a series of works related with spatiotemporal optical wavepacket, more specifically, the spatiotemporal optical vortices (STOV) wavepacket[20,21]. These works on STOV have led to a new research direction and have stimulated intense research interests in spatiotemporal optics[22-24], a rapidly developing field within which scientists study the light fields with a spatiotemporal coupled distribution, enabling light with new physical properties such as propagating in a controllable group velocity[25], achieving negative refraction at interfaces[26], and realizing the spatiotemporal mode-locking in lasers[27].

Recent progress in spatiotemporal optics leads us to the question: is it possible to build a new photonic quantum emulator in the form of spatiotemporal light? Since the evolution of a spatiotemporal light field is governed by the paraxial wave equation, which has the exact same mathematical form as the potential-free Schrödinger equation, the spatiotemporal light can mimic the wavefunction in the Schrödinger equation and can be exploited as a photonic quantum emulator for studying wavefunction behavior in quantum mechanics.

In this work, we demonstrate a new photonic quantum emulator by spatiotemporal light with spherical harmonic symmetry. This study has advanced previous studies in spatiotemporal optics as we, for the first time, manage to implement the full space-time, three-dimensional control of the light field. The generated spatiotemporal wavepacket is a special case of localized waves in optics[28] and it is a series of eigen-state solutions to the paraxial wave equation,

$$i\frac{\partial \psi}{\partial Z} = -\left(\frac{\partial^2}{\partial X^2} + \frac{\partial^2}{\partial Y^2} + \frac{\partial^2}{\partial T^2}\right)\psi. \tag{1}$$

Here, X, Y, and T are normalized coordinates and the choice of the signs in the equation indicates the optical medium has an anomalous group velocity dispersion ($\beta_2 < 0$). Mathematically, apart from the scaling factor, Equation (1) has the exact form as the potential-free Schrödinger equation, that is,

$$i\hbar\frac{\partial \psi}{\partial t} = -\frac{\hbar^2}{2m}\left(\frac{\partial^2}{\partial x^2} + \frac{\partial^2}{\partial y^2} + \frac{\partial^2}{\partial z^2}\right)\psi, \tag{2}$$

where only the time and the longitudinal coordinate are swapped in those two equations. Therefore, the localized waves solution to Eqn. (1) with two available quantum numbers is also the eigen-state wavefunction to the potential-free Schrödinger equation. In optics, such wavepackets solution is non-diffracting and non-dispersive, meaning the wavepacket can propagate without spreading in both space and time, like particles. From quantum mechanical perspective of view, the wavepacket resembles the electron in a "potential-free" hydrogen atom or a hydrogen atom without any Coulomb force between the electron and the nucleus. It is noteworthy that although previous works have studied other optical localized waves such as Bessel X-wave[29], Bessel O-wave[30], and many others[31,32], none of these light fields has spherical harmonic symmetry or quantum-like analogy, which is mainly due to the lack of effective technique or instrument for manipulating the light field three-dimensionally.

In this work, we manage to generate a series of three-dimensional spatiotemporal localized wavepackets with spherical harmonic symmetry as a new type of photonic quantum emulator. These wavepackets are generated, for the first time in the laboratory, using the combination of spatiotemporal hologram[33] and conformal mapping[34] techniques. We generate and measure these spherical harmonic localized wavepackets with different combination of quantum numbers $l$ and $m$. All of them have a three-dimensional distribution that is the eigen-state solution to the potential-free Schrödinger equation. The optical non-varying and self-healing properties of these spherical harmonic localized wavepackets are experimentally verified, confirming their eigen-state feature. These spatiotemporal wavepackets forms a new class of photonic quantum emulator, enabling new quantum mechanical studies in the dynamics of eigen-state wavefunction. The localized feature of these spatiotemporal wavepackets can also find new opportunities in optical applications such as bioimaging[35,36], optical tweezers[37], optical communication[38], and laser machining[39].

## Results and Discussions

The electric field of an optical wavepacket can be described as the product of an envelope function and the carrier oscillation term $E(x,y,t;z) = \psi(x,y,t;z) \cdot \exp i(\omega_0 t - kz)$. For a spherical harmonic localized wavepacket, its envelope function $\psi(x,y,t;z)$ is the solution to the paraxial wave equation[28]. In a spherical coordinate, the envelope function with quantum numbers of $(l,m)$ can be expressed as

$$\psi_{(l,m)}(X,Y,T,Z) = \psi_0 j_l(\alpha R) P_l^m(\cos\theta) \exp(im\phi) \exp(-i\alpha^2 Z). \qquad (3)$$

Noteworthy, Equation (1) denotes the solutions have exchangeability of the dimensionless spatiotemporal coordinates $(X,Y,T)$. In this paper, we study LW wavepacket with its polar axis pointing to the $T$-axis. The spiral phase term $\exp(im\phi)$ is then set in the X-Y plane. In Equation (3), $(R,\theta,\phi)$ are normalized spherical coordinates with $R = \sqrt{X^2 + Y^2 + T^2}$, $\theta = \cos^{-1}\frac{T}{R}$ and $\phi = \tan^{-1}\frac{Y}{X}$, $j_l$ is the spherical Bessel function of the first kind, $P_l^m$ is the associated Legendre polynomials, and $\alpha$ is the propagated constant. $l$ and $m$ are integer numbers with $|m| \leq l$. For a spherical harmonic localized wavepacket, $l$ stands for the azimuthal number and $m$ stands for the topological charge of longitudinal OAM. In quantum mechanics, $l$ stands for the quantum number of total orbital angular momentum. $m$ is related with the projection of the orbital angular momentum in the polar direction, also known as the "magnetic" quantum number. Since this solution is to the potential-free Schrödinger equation, the prime quantum number $n$ is absent in this expression.

The generation of three-dimensional wavepackets requires an ability to directly sculpt wavepackets in the full spatiotemporal domain (X, Y, and T), which is difficult to implement in the laboratory, especially for wavepackets with a sub-picosecond pulse duration. Recently, the generation of optical toroidal pulse achieved a three-dimensional spatiotemporal optical field with the aid of conformal mapping technique for implementing a Cartesian to log-polar transformation[40]. The conformal mapping technique offers an alternative route for creating 3D wavepackets with rotational symmetry along the propagated axis, especially when the technique is combined with the newly developed spatiotemporal holographic pulse shaping approach. In a conformal mapping process, the spatial coordinate is transformed from Cartesian coordinates $(x,y)$ to log-polar coordinates $(u,v)$, requiring $u = b \cdot \exp\left(-\frac{x}{a}\right) \cdot \cos\left(\frac{y}{a}\right)$ and $v = b \cdot \exp\left(-\frac{x}{a}\right) \cdot \sin\left(\frac{y}{a}\right)$. The transformation is realized by an afocal system consisting of two phase masks. The first phase mask is[34]

$$\phi_1(x,y) = \frac{k}{d}\left[-ab \cdot \exp\left(-\frac{x}{a}\right) \cdot \cos\left(\frac{y}{a}\right) - \frac{x^2+y^2}{2}\right], \tag{3}$$

where $k$ is the wave number, $d$ is the distance between two phase masks, $a$ and b are the position and size parameter for the conformal mapping process[34, 40]. The first phase mask $\phi_1(x,y)$ can realize the transformation and the second phase mask $\phi_2(u,v)$ corrects the phase distortion by

$$\phi_2(u,v) = \frac{k}{d}\left[-au \cdot \ln\frac{\sqrt{u^2+v^2}}{b} + av \cdot \tan\frac{v}{u} + au - \frac{u^2+v^2}{2}\right]. \tag{4}$$

This conformal mapping realizes the Cartesian-to-log-polar coordinate transformation. Conversely, a ring-like beam can be mapped to a line when it is back-propagated through the same conformal mapping system. For the spherical harmonic localized wavepackets $\psi_{(l,m)}$ described in Eqn. (1), its back-propagated field prior to the conformal mapping can be expressed in the Cartesian coordinates as

$$\Psi_{(l,m)}(x,y,\tau) = -\frac{\rho}{a^2}\exp\left(\frac{\rho^2+\tau^2}{w^2}\right)j_l\left(\alpha\sqrt{\rho^2+\tau^2}\right)P_l^m\left(\frac{\tau}{\sqrt{\rho^2+\tau^2}}\right)\exp\left(im\frac{y}{a}\right), \tag{5}$$

where $\rho = b \cdot \exp(-x/a)$, $\tau$ is the retarded time. If we examine the spherical harmonic localized wavepacket solution in Eqn. (1), the wavepacket prior to the conformal mapping process is a two-dimensional (2D) spatiotemporal optical field with an additional linear phase along the $y$-direction. Such a field $\Psi_{(l,m)}$ can be generated by the spatiotemporal holographic pulse shaping technique. Figure 1(a), (b), and (c) illustrate the back-propagated field prior to the conformal mapping for the spherical harmonic localized wavepackets with quantum numbers $(l,m)$ of (1,1), (2,1) and (3,1). The right panels shown in the figures show the generated spherical harmonic localized wavepackets after conformal mapping.

Figure 2 depicts the experimental setup for generating and measuring the spherical harmonic localized wavepackets, which constitutes three parts: the spatiotemporal holographic pulse shaping stage, the conformal mapping stage, and the measurement stage. The input laser is firstly divided by a beamsplitter (BS1) into the "signal" pulse (red beam) and the "probe" pulse (blue beam). In the "signal" pulse arm, the "signal" pulse is tailored into $\Psi_{(l,m)}(x,y,\tau)$ after passing the spatiotemporal holographic pulse shaper (Grating, CL1, and SLM1). The shaper provides the spatiotemporal complex amplitude modulation for input pulses through spatiotemporal holography[33]. Then, a one-dimensional telescopic system (CL2 and CL3) expands the beam $\Psi_{(l,m)}(x,y,\tau)$ uniformly in the y-direction with a magnification factor of 14. In the second stage, the conformal mapping system (SLM2 with a mapping phase of $\phi_1$ and SLM3 with a correction phase of $\phi_2$) generates the desired spherical harmonic localized wavepacket $\psi_{(l,m)}$. Here, SLM3 also adds the spatial spiral phase to wavepacket for creating the desired topological charge of m. In the "probe" arm, the "probe" pulse is compressed to 100 fs and then enters the measurement system. The measurement system scans the relative time delay between the "signal" and the "probe" for recording the time-scanned interferometric patterns between them. Finally, the three-dimensional wavepacket retrieval algorithm is applied to reconstruct the generated spherical harmonic localized wave wavepackets[42, 43].

Figure 3 plots the reconstructed results for spherical harmonic localized wave wavepackets with quantum numbers $(l,m)$ of (1,1), (2,1), and (3,1). Figure 3(a) plots the back-propagated field of $\Psi_{(1,1)}$ for its spatiotemporal intensity and phase distribution. Figure 3(b) shows the 3D intensity isosurface of the generated spherical harmonic localized wavepacket after the conformal mapping process measured $\psi_{(1,1)}$ and its phase distribution at $\tau = 0$. The phase plot indicates the localized wave wavepacket has a topological charge of $m = +1$. The reconstructed 3D intensity isosurface of $\psi_{(2,1)}$ and $\psi_{(3,1)}$ are shown in Fig. 3(c) and (d).

Being a solution to Equation (1), the spherical harmonic localized wavepacket can propagate without changing its form in an anomalous dispersive medium. In the experiment, we demonstrate this non-varying feature by propagating the generated spherical harmonic localized wavepackets $\psi_{(1,1)}$ in a "virtual" dispersive medium with a group velocity dispersion coefficient $\beta_2$ of -105 fs²/mm. The "virtual" dispersion is controlled by SLM1 in the setup. The corresponding dispersion length for the wavepacket $L_{\text{dis}} = \tau_0^2/|\beta_2|$ is 231 mm, same as the diffraction length $L_{\text{diff}} = k_0 w_0^2$. In the calculation, the experimental parameters are $w_0 = 192$ μm and $\tau_0 = 156$ fs. Figure 4 shows the propagation dynamics of the experimental generated $\psi_{(1,1)}$ in the "virtual" anomalous dispersive medium for a propagation distance of $0L_{\text{diff}}$, $2L_{\text{diff}}$, and $4L_{\text{diff}}$. The three-dimensional distribution of the wavepacket has no significant change over 4 diffraction length. In the same figure, we plot the theoretical propagation dynamics of the wavepacket. Our experimental results have a good agreement with the numerical simulation results.

Another intriguing feature of the spherical harmonic localized wavepacket is its self-healing property as it can recover its original field distribution after being partially blocked by an obstacle. To verify this property, we blocked the generated $\psi_{(1,1)}$ wavepacket using a fiber lead with a width of 250 μm. After being partially blocked (see Fig. 5(a)), the wavepacket can self-heal after it undergoes a dispersive propagation. Figure 5(b) and (c) plot the spatiotemporal intensity distribution of $\psi_{(1,1)}$ after it propagates $1L_{\text{diff}}$ and $2L_{\text{diff}}$. The results clearly show the self-healing process is occurring in the full spatiotemporal domain and the resulting wavepacket has no significant change after its central lobe being partially blocked.

## Conclusions

In summary, we demonstrate a new class of photonic quantum emulator enabled by spatiotemporal localized wavepackets with spherical harmonic symmetry. Being an eigen-state solution to the paraxial equation, the spherical harmonic localized wavepackets have the same distribution of the eigen-state solution to the potential-free Schrödinger equation. In the experiment, the wavepackets are generated by spatiotemporal holographic shaping and conformal mapping techniques. The localized feature of these wavepackets is also experimentally verified by showing its non-varying and self-healing properties.

The use of conformal mapping and spatiotemporal holography offers a new route for sculpting spatiotemporal light fields with three dimensional features, leading to the possibilities of generating complicated spatiotemporal light. Besides their potential applications in bio-imaging, optical tweezer, and optical communication, and laser machining, these spherical harmonic localized wavepackets can serve as a new class of photonic quantum emulator for investigating the behavior of eigen-state wavefunction, emulating light-and-matter interactions, and finding applications in quantum optics.

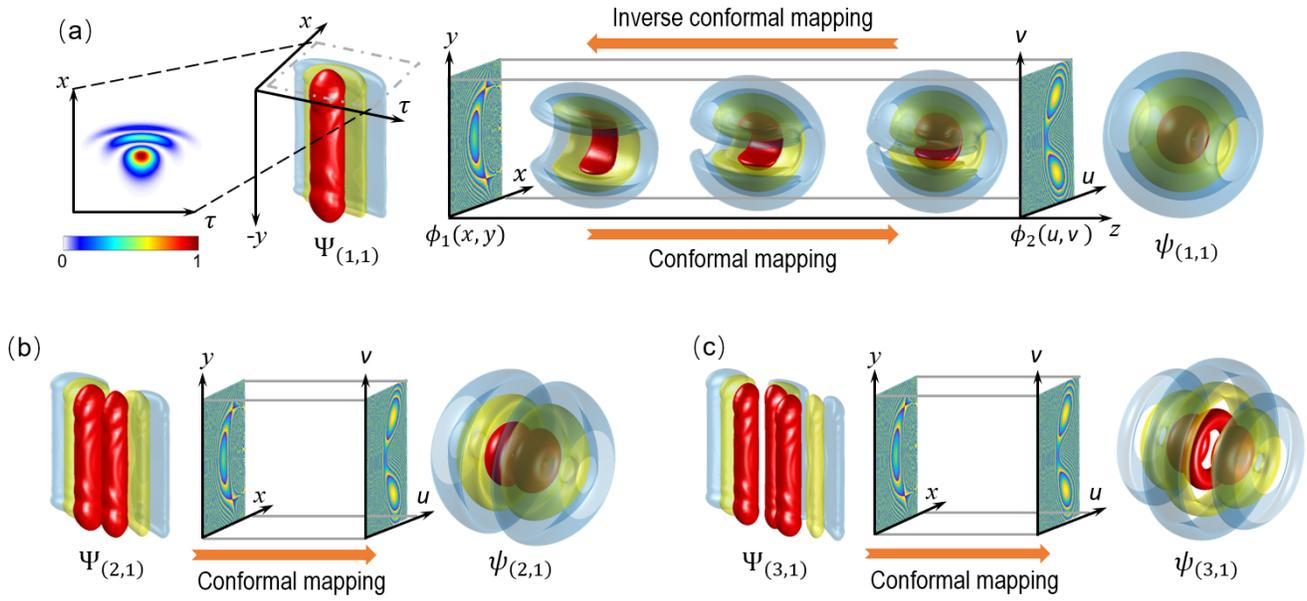

**Figure 1. Inverse conformal mapping and conformal mapping for designing and generating spherical harmonic localized wavepackets $\psi_{(l,m)}$ with quantum numbers of (*l,m*).** (a) Using the back-propagation method to design the inversely conformal mapped spatiotemporal light field $\Psi_{(l,m)}$. In the forward direction, the field is transformed into the spherical harmonic localized wavepacket $\psi_{(1,1)}$ with quantum numbers of (1,1). (b) Generation of spherical harmonic localized wavepacket with quantum numbers of (2,1). (c) Generation of spherical harmonic localized wavepacket with quantum numbers of (3,1).

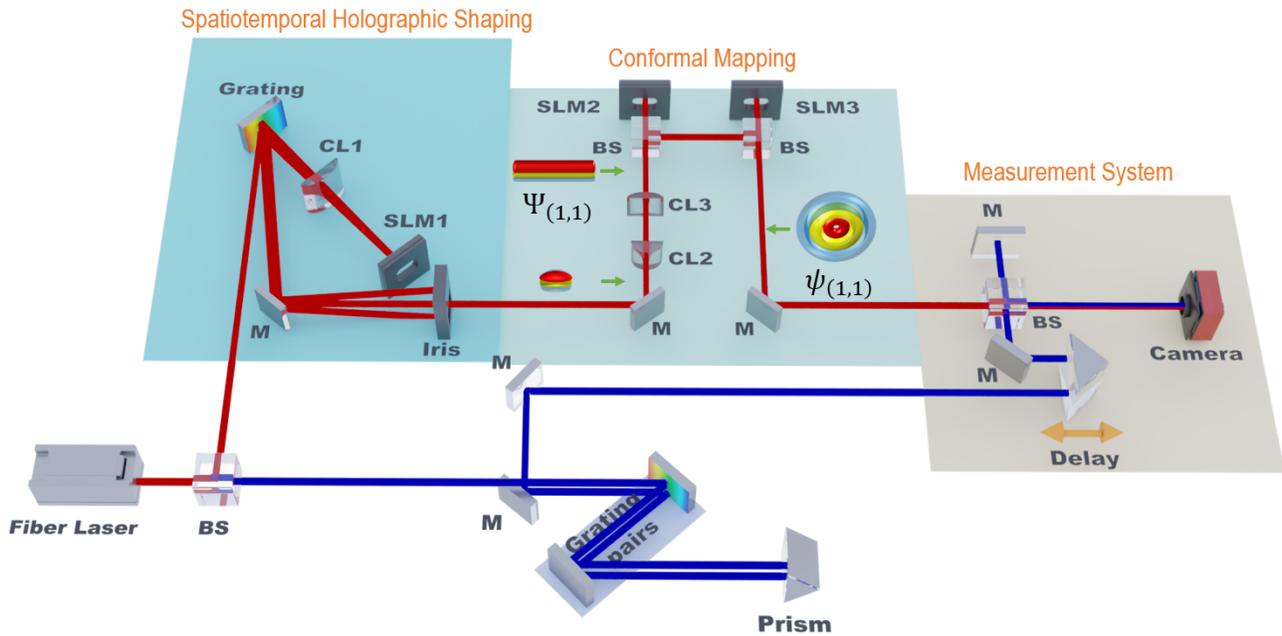

**Figure 2. Schematic of the experimental setup for generating and measuring spherical harmonic localized wavepackets.** The optical setup has a Mach-Zehnder interferometric schematic. The laser input is firstly split into two copies, the "signal" pulse (red beam path) and the "probe" pulse (blue beam path) by a beamsplitter (BS1). The "signal" pulse is modulated into spherical harmonic localized wavepackets by the cascaded use of spatiotemporal holographic shaper (Grating, CL1, SLM1, and Iris, see [20]) and conformal mapping system (CL2, CL3, SLM2, and SLM3). The "probe" pulse (blue optical path in the figure) is temporally compressed and is used to interfere with the generated localized waves "signal" wavepacket in the measurement system for retrieving the three-dimensional profile of the generated "signal", the spherical harmonic localized wavepacket.

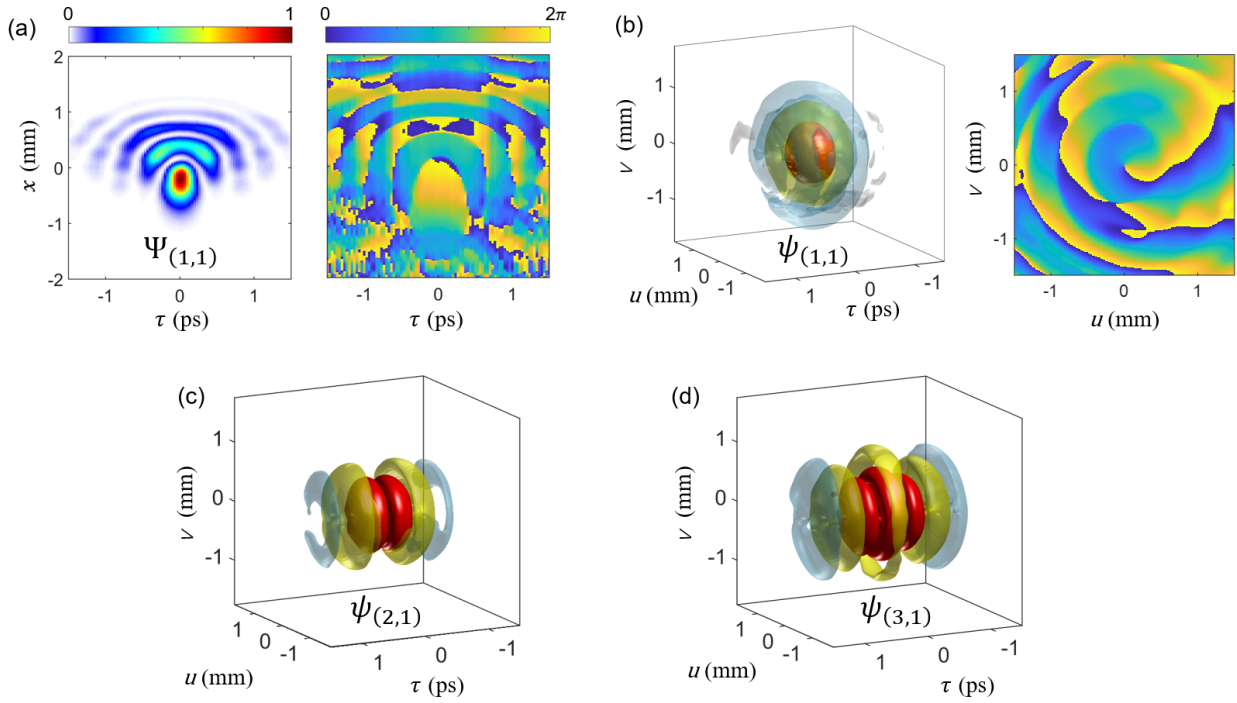

**Figure 3. Experimental measurement of spherical harmonic localized wavepackets with different combination of quantum numbers.** (a) Spatiotemporal intensity and phase distributions of the wavepacket $\Psi_{(1,1)}$ prior to the conformal mapping. (b) 3D iso-intensity surface plot and the spatial phase for $\psi_{(1,1)}$. The isosurface value is set at 1% of the peak intensity. Different colors are used to indicate different ring-like lobes of $\psi_{(1,1)}$. The phase profile corresponds to the spatial phase at $\tau = 0$. (c) 3D iso-intensity surface plots for $\psi_{(2,1)}$. (d) 3D iso-intensity surface plots for $\psi_{(3,1)}$.

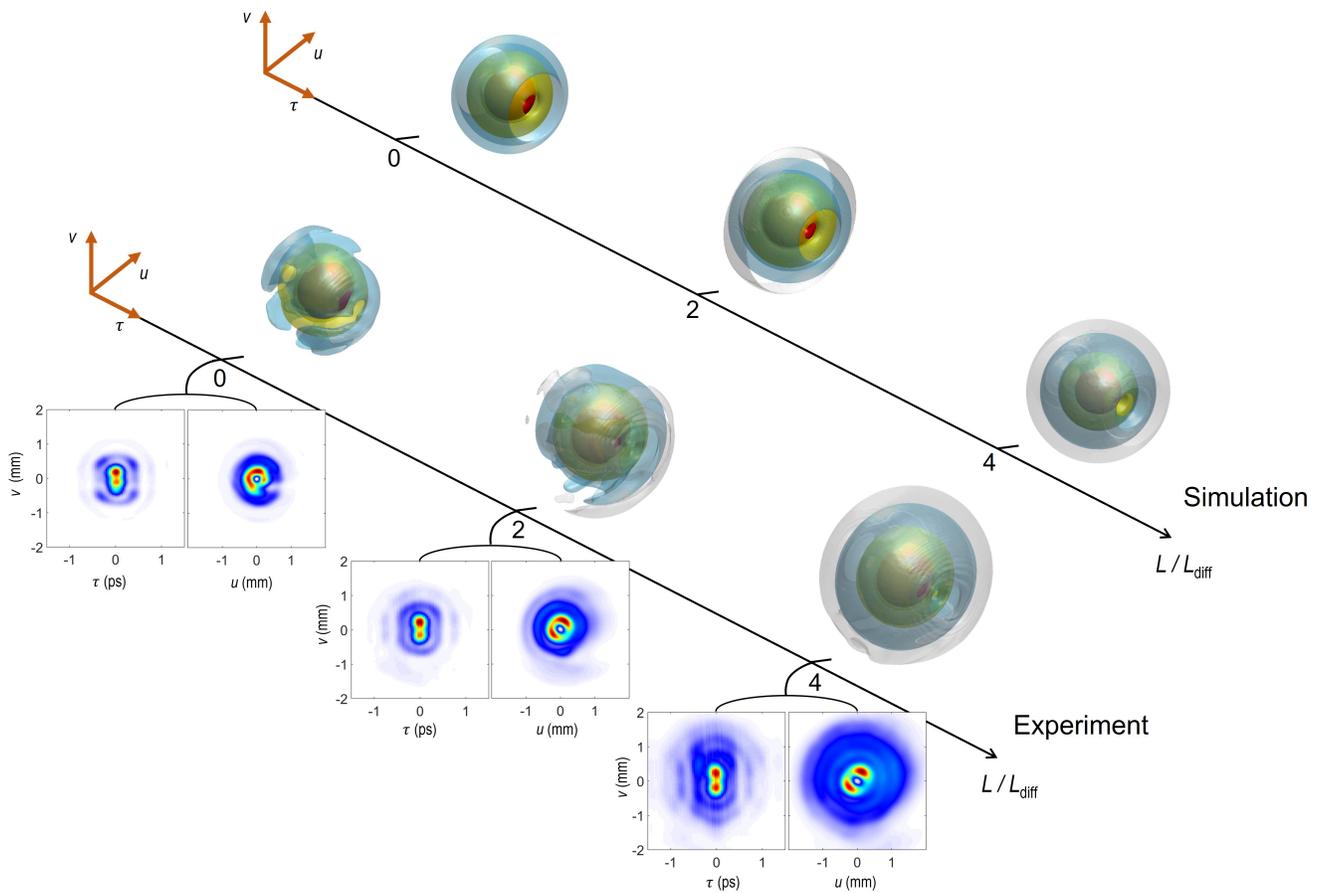

**Figure 4. Non-varying property of spherical harmonic localized wavepacket.** The spherical harmonic localized wavepacket $\psi_{(1,1)}$ with quantum numbers of $(1,1)$ propagates in a "virtual" dispersive medium for verifying its non-varying feature. The propagation dynamics of the wavepacket is shown at $0L_{\text{diff}}$, $2L_{\text{diff}}$ and $4L_{\text{diff}}$. The experimental measurement results well agree with the simulation results.

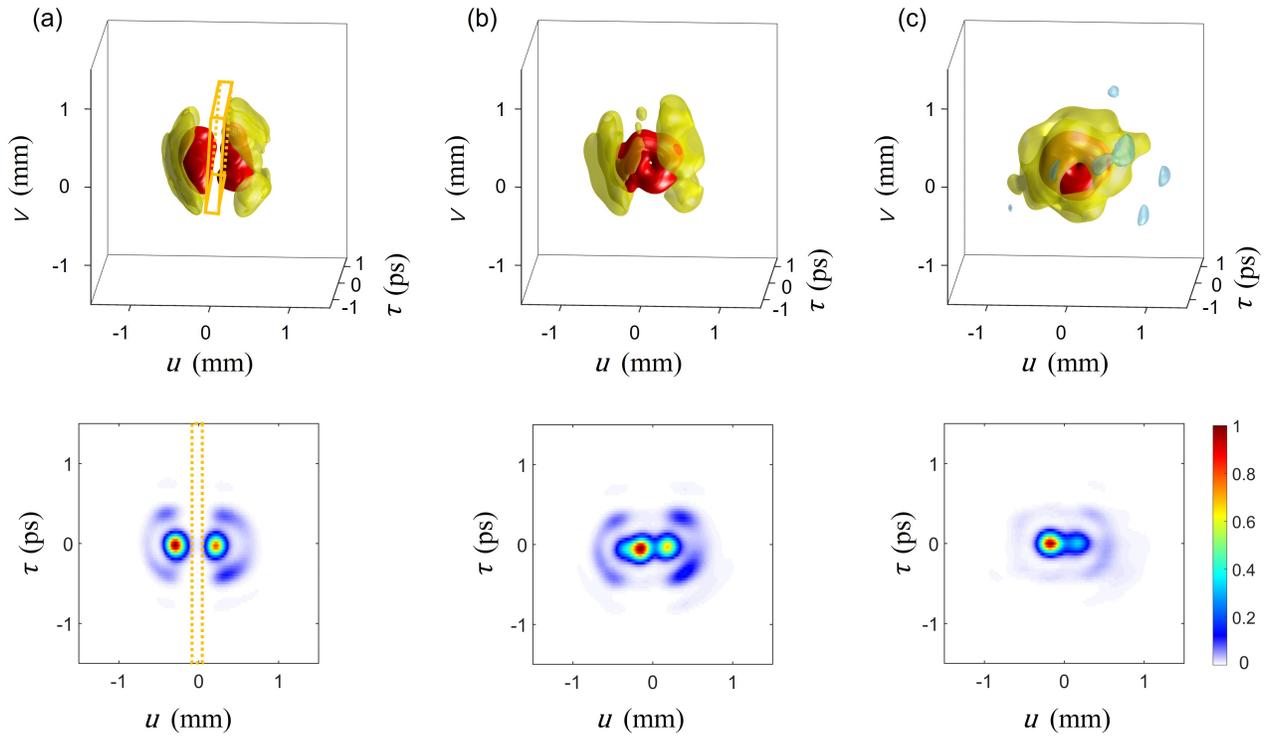

**Figure 5. Self-healing property of spherical harmonic localized wavepacket.** After a spatial blockage from a 250-μm-wide fiber lead, the wavepacket can self-heal itself three-dimensionally after it propagates in a dispersive medium. (a) 3D intensity profile of the experimentally generated $\psi_{(1,1)}$ wavepacket after a spatial blockage. (b) Profile of the wavepacket after $1L_{\text{diff}}$. (c) Profile of the wavepacket after $2L_{\text{diff}}$.

**Availability of data and materials**

The datasets used and/or analysed during the current study are available from the corresponding author on reasonable request.

**Competing interests**

The authors declare no competing interest.

**Funding**

We acknowledge financial support from National Natural Science Foundation of China (NSFC) [Grant Nos. 12434012 (Q.Z.) and 12474336 (Q.C.)], the Shanghai Science and Technology Committee [Grant Nos. 24JD1402600 (Q.Z.) and 24QA2705800 (Q.C.)], National Research Foundation of Korea (NRF) funded by the Korea government (MSIT) [Grant No. 2022R1A2C1091890], and Global - Learning & Academic research institution for Master's·PhD students, and Postdocs (LAMP) Program of the National Research Foundation of Korea(NRF) grant funded by the Ministry of Education [No. RS-2023-00301938]. Q.Z. also acknowledges support by the Key Project of Westlake Institute for Optoelectronics [Grant No. 2023GD007].

**Authors' contributions**

QC and NZ implemented the experiment and analyzed all the data. QC, AC, and QZ conceptualized and designed the project. AC and QZ supervised the project. QC and NZ wrote the original draft. QC, AC, and QZ reviewed and finalized the manuscript. All authors read and approved the final manuscript.